\documentclass{emulateapj}

\usepackage{amssymb,gensymb,graphics,textcomp}

\newcommand{\OI}{[\ion{O}{1}]}

\newcommand{\Msun}{$M_{\odot}$}

\newcommand{\OIII}{[\ion{O}{3}]}
\newcommand{\SII}{[\ion{S}{2}]}
\newcommand{\NII}{[\ion{N}{2}]}

\newcommand{\Ha}{H$\alpha\,$}
\newcommand{\Hb}{H$\beta\,$}

\newcommand{\HII}{\ion{H}{2}}

\shorttitle{}
\shortauthors{Rich et al.}

\begin{document}

\title{NGC~839: Shocks in an M82-like Superwind}

\author{J. A. Rich \altaffilmark{1}, M. A. Dopita\altaffilmark{1}\altaffilmark{2}, L. J. Kewley\altaffilmark{1} \& D. S. N. Rupke\altaffilmark{1} }
\email{jrich@ifa.hawaii.edu}
\altaffiltext{1}{Institute for Astronomy, University of Hawaii, 2680 Woodlawn Drive, Honolulu, HI 96822}
\altaffiltext{2}{Research School of Astronomy and Astrophysics, Australian National University, Cotter Road, Weston ACT 2611, Australia }

\date{\today}

\begin{abstract}
We present observations of NGC~839 made with the Wide Field Spectrograph on the ANU 2.3m telescope. Our data cover a region $25\arcsec \times 60$\arcsec~at a spatial resolution of $\sim1\farcs5$ The long axis of the field is aligned with the superwind  we have discovered in this starburst galaxy. The data cover the range of 3700-7000\AA, with a spectral resolution $R\sim7000$ in the red, and $R\sim3000$ in the blue. We find that the stellar component of the galaxy is strongly dominated by a fast rotating intermediate-age ($\sim 400$Myr) A-type stellar population, while the gas is concentrated in a bi-conical polar funnel. We have generated flux distributions, emission line ratio diagnostics and velocity maps in both emission and absorption components. We interpret these in the context of a new grid of low-velocity shock models appropriate for galactic-scale outflows. These models are remarkably well fit to the data, providing for the first time model diagnostics for shocks in superwinds and strongly suggesting that shock excitation is largely responsible for the extended LINER emission in the outflowing gas in NGC 839. Our work may have important implications both for extended LINER emission seen in other galaxies, and in the interpretation of objects with ``composite" spectra. Finally, we present a scenario for the formation of E+A galaxies based upon our observations of NGC~839, and its relation to M82.\\
\end{abstract}



\section{Introduction}
The physical nature of the low-ionization narrow emission-line regions (LINERs) in galaxies has been an area of intense study and debate since their definition by \citet{Heckman80}. Work by ~\citet{Ho97b},~\citet{Kewley06} and \citet{CidFernandes10} inter alia suggests that this class of object represents the most common type of non-starbursting galaxy. LINERs, as their name suggests, are defined by their relatively strong low-ionization state lines (e.g., \OI,  \NII and \SII). They are found both in the nuclei of giant elliptical galaxies, or, as in the case of NGC~839, in lower mass gas-rich star-forming systems. 

The excitation mechanisms proposed to explain the LINER phenomenon have included excitation by weak or low-luminosity active galactic nuclei (AGNs; \citet{Ferland83,Halpern83,Ho99,Kewley06,Ho09,Eracleous10}, young and/or post-asymptotic giant branch (AGB) stars \citep{Terlevich85,Trinchieri91,Shields92,Binette94,Stasinska08,Sarzi10}, shocks \citep{Heckman80,Dopita96,Kewley01a,Lipari04} and cooling flows \citep{Voit97}. Indeed it is likely that in many galaxies, and in different regions of the same galaxy, more than one of these mechanisms may be at work. Recent studies of mostly passive, early-type galaxies \citep{Sarzi10, Annibali10, Eracleous10} present evidence both for and against weak AGNs, post-AGB (PAGB) stars and shocks in various systems. Although these authors find some evidence for excitation due to PAGB stars or weak AGNs, neither mechanism provides a wholly satisfactory answer in all cases.  The most compelling evidence in support of the PAGB mechanism is the finding that (in elliptical galaxies) the emission-line flux correlates very well with the host galaxy stellar luminosity within the emission-line region \citep{Macchetto96}, and that the emission-line flux distribution follows that of the stellar continuum \citep{Sarzi06,Sarzi10}. However, \citet{Binette94} and \citet{CidFernandes09} demonstrated that photoionization by PAGB stars can only explain LINERs with relatively weak emission lines. The \Ha\ equivalent width produced by an 8 Gyr old and a 13 Gyr passively-evolving stellar population amounts to 0.6 \AA~and 1.7 \AA, respectively.  Stronger emission lines require far more ionizing photons than old stars can provide.

However, studies of more active galaxies and mergers, notably those with winds, find strong evidence for the dominance of emission by shock excitation \citep{Veilleux02,Veilleux03,Lipari04,Monreal10,Sharp10}. The line ratios observed in the ionized gas are typically consistent with LINER-like emission, although this class of emission can extend far from the nucleus and disk of the host galaxy. In this case, the gas may lie far from the nucleus and is not ionized directly by nuclear activity. In more distant galaxies, the entrance aperture of the spectrograph may capture both the circum-nuclear and the extra-nuclear components, making the classification still more difficult.

In order to separate the various spatial components of the ionized plasma, and to disentangle the potentially complex kinematics and diverse array of power sources in galaxies, we need to employ integral field units (IFUs). These are increasingly providing insight into complex galactic systems. Several wider-field IFUs have been commissioned and used, notably the SAURON instrument \citep{dezeeuw02}, which has been extensively used to survey early-type galaxies. The SINS survey employed the SINFONI instrument \citep{Genzel08} to examine high-$z$ systems. The VIMOS IFU \citep{Arribas08} has been used for a survey of Ultraluminous Infrared galaxies, and the PINGS survey of nearby disk galaxies \citep{Rosales10}. Finally, the SPIRAL instrument ~\citep{Sharp10} has been employed to look at galactic winds. In this paper, we will describe data obtained of the superwind galaxy NGC~839 using the Wide Field Spectrograph (WiFeS). This instrument is described in detail in \citet{Dopita07,Dopita10}.

NGC~839 is a nearly edge-on disk galaxy ($I=67$\textdegree, P.A.$=350$\textdegree) in a Hickson Compact Group (HCG 16): a dense, interacting group of seven spirals, six of which are actively star forming \citep{Rubin91,Ribeiro96}. NGC 839 may well be a recent addition to the group, more kinematically disturbed and showing signs of an ongoing or recent interaction \citep{Rubin91,Mendes98,deCarvalho99}. \citet{Belsole03} have detected diffuse, hot X-ray emitting gas encompassing the whole group, evidence that it forms a bound group even though it has no bright early-type galaxies. 

Multiwavelength observations of NGC~839 indicate both ongoing star formation and LINER-like line ratios. \citet{Ribeiro96} used the \NII/\Ha ratio to classify NGC~839 as a starburst, though their measured values actually fall in the composite starburst/AGNs region when the \citet{Kewley06} scheme is applied. In addition, \citet{deCarvalho99} have employed the observed \NII/\Ha and \SII/\Ha ratios to classify NGC~839 as a LINER + Seyfert double nucleus.  X-ray spectra of the galaxy indicate the presence of an active starburst coupled with slightly higher than solar metallicities and may indicate the presence of a low luminosity, obscured AGN~\citep{Turner01}. ~\citet{GonzalezMartin06} however find no evidence for an unresolved point source in hard X-ray images and classify NGC~839 as a starburst galaxy using X-ray morphology.

\begin{figure}
\centering
{\includegraphics[scale=0.7]{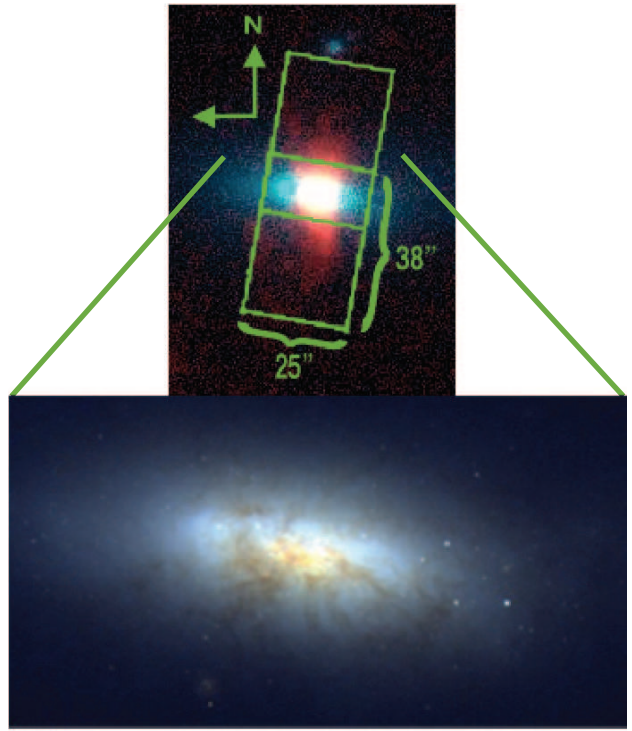}}\label{fig1}
\caption{Composite $R$+\Ha (green + red) image of NGC 839 from~\cite{Meurer06} and the archival \emph{Hubble Space Telescope} Advanced Camera for Surveys (\emph{HST} ACS) composite $B$+$R$+$I$ blow-up of the disk region.  The two WiFeS pointings are indicated by the green rectangles overlaid on the SINGG image.}
\end{figure}

At IR-wavelengths, NGC~839 is clearly classified as a Luminous Infrared Galaxy (LIRG) with $L_{IR}\sim10^{11}L_{\odot}$ \citep{Armus09}. This luminosity derives from a nuclear starburst, which is probably driving an M~82-like galactic wind, evidenced by the very extended polar funnels of \Ha emission seen by the Survey for Ionization in Neutral Gas Galaxies (SINGG) \Ha images (\citet{Meurer06}; see Figure \ref{fig1}).

The observations presented in this paper are the first results from the WiFeS GOALS survey. The WiFeS GOALS survey is a sub-sample drawn from the larger Great Observatory All-Sky LIRG Survey (GOALS) sample. We discuss the observations and data reduction in Section \ref{observations}. In Section \ref{results}, we present the results of our observations and analysis and present spectra, rotation curves, and line ratio maps. In Section \ref{analysis}, we compare new photoionization models and low-velocity shock models to the observations, and compare the derived properties of NGC~839 to the prototypical superwind galaxy M82. We conclude that the balance of evidence is in favor of the superwind of NGC 839 being shock-excited. Finally, in Section \ref{conclusions}, we discuss the relationship of NGC~839 to ``composite '' (or transition) galaxies and suggest extended ``composite " or LINER emission may be fueled by shock emission. We also suggest that NGC~839 is one possible ``missing link'' to E+A galaxies and suggest an evolutionary scenario to explain how this class of objects might have arisen.

Throughout this paper, we adopt the cosmological parameters $H_{0}$=70.5~km~s$^{-1}$Mpc$^{-1}$, $\Omega_{\mathrm{V}}$=0.73, and $\Omega_{\mathrm{M}}$=0.27, based on the 5 year WMAP results \citep{Hinshaw09}. With these parameters, and taking into effect both the Virgo and Shapley infall, the adopted distance to the HCG 16 is $55\pm1$~Mpc (NASA Extragalactic Database, NED). This gives a spatial scale for NGC~839 of 260~pc arcsec$^{-1}$. \newline

\section{Observations, Data Reduction and Analysis} \label{observations}
\subsection{Observations}
Our data of NGC 839 were taken using the WiFeS at the Mount Stromlo and Siding Spring Observatory 2.3 m telescope.  WiFeS is a new, dual beam, image-slicing IFU described in detail by \citet{Dopita07,Dopita10}. The instrument is mounted in a stationary position at a Nasmyth focus to provide excellent stability. The IFU field consists of $25 \times 1$\arcsec~wide slitlets, each of which is  38\arcsec~long. The spatial pixel is $0''.5$ along the slitlet axis and $1''.0$ in the spectral direction. In the blue, the spectra cover the range of 3700-5700 \AA, at a spectral resolution of $R\sim3000$ ($\approx100$ km s$^{-1}$). In the red, we cover 5700-7000 \AA\ at a resolution of $R\sim7000$ ($\approx45$ km s$^{-1}$). Thus, the data have sufficient spectral resolution in the red to allow detailed dynamical studies, while the total wavelength coverage is sufficient to support excitation and chemical abundance analyses.

We observed NGC 839 on 2009 August 18, about 4 months after the instrument had been commissioned. Two pointings were made at a position angle (P.A.) of 350\textdegree~(see Figure \ref{fig1}), so that the slitlets were aligned with the minor axis of NGC 839. This allows us to best probe the previously detected extended \Ha emission which occurs in two funnels both above and below the disk~\citep{Mendes98,Meurer06}. Each pointing consisted of three 800s exposures giving a total integration time of 2400s for the extended emission, and 4800s in the overlapping region centered on the disk. The seeing was $\sim1.2$\arcsec~during the course of our exposures. Sky subtraction was carried out using a sky spectrum drawn from relatively clean sky regions at the ends of each data cube. 

\subsection{Data Reduction}
The data were reduced and flux calibrated using the WiFeS pipeline, briefly described in \citet{Dopita10}, which uses IRAF routines adapted primarily from the Gemini NIFS data reduction package. 

The bias subtraction is somewhat complicated by the use of quad-readout to  decrease chip read-out time as well as due to a slight slope and instability in the bias across each region of the chip. Bias frames are taken immediately before and after each set of observations and a two-dimensional fit of the surface is subtracted from the temporally nearest object data in order to avoid adding additional noise to the data. Any resulting residual is accounted for with a fit to unexposed regions of the detector.

Quartz lamp flats are used to account for the response curve of the chip and twilight sky flats are used to correct for illumination variation along each of the slitlets.  Spatial calibration is carried out by placing a thin wire in the filter wheel and illuminating the slitlet array with a continuum lamp. This procedure defines the center of each slitlet. The individual spectra have no spatial distortion because the camera corrects the small amount of distortion introduced by the spectrograph. Thus only low-order mapping of the slitlets is required. 

Wavelength calibration is performed using a CuAr arc lamp to provide sufficient lines in both the blue and red arms of the camera. Arc lamp data were taken in between sets of object exposures.  Each of the 25 slitlets is then rectified by the pipeline into a full data cube (one for each arm) sampled on a common wavelength scale.  The resulting data cubes from each exposure were flux and telluric calibrated using observations of the white dwarf Feige 110 and corrected for the effect of atmospheric dispersion. 

Cosmic-ray removal was performed with the "dcr" routine \citep{Pych04}. The final reduced and flux-calibrated data cubes were binned by 2 pixels along the slit in order to increase signal-to-noise ratio and to produce square spatial elements $1\arcsec \times 1$\arcsec, corresponding to $\sim260$ pc at the distance of the galaxy. 

We analyzed each spectrum using an automated fitting routine written in IDL, UHSPECFIT, which is based on the code created by \citet{Zahid10} and is also employed by \citet{Rupke10}. Our routine fits and subtracts a stellar continuum from each spectrum using population synthesis models from \citet{Gonzalez05} and an IDL routine which fits a linear combination of stellar templates to a galaxy spectrum using the method of \citet{Moustakas06}. Although our population synthesis models include stellar absorption features, it was necessary to fit excess absorption in \Hb or \Ha in some spectra to ensure a reliable emission line flux estimates. We believe this is due to the inadequacy of the stellar models. Lines in the resulting emission spectra are fit using a one or two-component Gaussian, depending on the goodness of fit determined by the routine. All of the emission lines are fit simultaneously using the same Gaussian component or components. Resulting fits were given a cursory inspection by eye to ensure the routine did not fail. Both continuum and emission lines were fit using the MPFIT package, which performs a least-squares analysis using the Levenberg--Marquardt algorithm \citep{Markwardt09}.

\begin{figure}
\centering
{\includegraphics[scale=0.325]{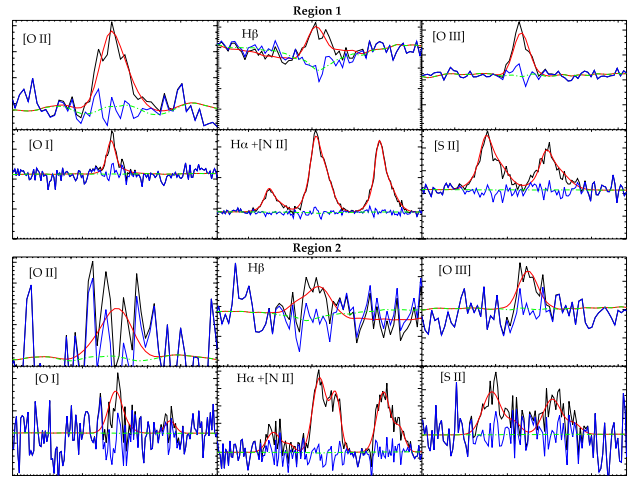}}\label{fig2}
\caption{Example of the line and continuum fitting procedure described in the text applied to two spatial pixels extracted from the image cube. Note the clear line splitting seen in the second example. Each box has been scaled to a total velocity range of 3000 km~s$^{-1}$ in the $x$-direction and from an intensity of zero up to the peak intensity in the  $y$-direction. The raw data are in black, the continuum fit is in green, and continuum+emission fit is in red. The lower blue curve gives the residuals to the fit.}
\end{figure}

The fitting routine provides a redshift, flux, and width for each Gaussian component in each spatial pixel. The blue and red spectra are fit simultaneously, while accounting for slight variations in wavelength zero point and instrumental resolution between the two arms as parameters in the fitting routine. The \Hb, \OIII\ 5007\AA, \OI\ 6300\AA, \NII\ 6583\AA, \Ha, and \SII\ 6717,6731\AA  ~emission lines were fit reliably across the disk and in the extended emission. Where there are two components, we consistently apply the "principal" label to the component with the strongest Ha emission line flux.

We apply a signal-to-noise cutoff of $5\sigma$ for accurate measurement of emission line fluxes. This sets an effective flux sensitivity of $\sim2.5 \times 10^{-17}$erg s$^{-1}$ cm$^{-2}$ in \OI, \NII, \SII\ and \Ha\ and $\sim1.5 \times 10^{-17}$erg s$^{-1}$ cm$^{-2}$ in \OIII\ and \Hb. In all, some 300 spatial pixels (spaxels) lie above the $>5\sigma$ cutoff in the strongest line (\NII\ or \Ha) out of a total of $\sim1500$ possible spatial elements in the final binned and mosaicked cube.  

An example of one of our fits is shown in Figure 2. In this example, the line splitting is clearly evident, as well as the underlying stellar continuum. In this example, all of the emission lines shown in Region 1 are above the signal-to-noise cutoff mentioned above, while in Region 2 only \Ha\ and \NII\ are above the cutoff, the rest of the emission line fluxes are thrown out for the purposes of our analysis. The cutoff is applied to each component in order to avoid contamination from a non-detected second component in weaker lines---in Figure 2, the \OI\ line in Region 1 is an example of a line where the primary component is above the cutoff while the second component is not.

\section{Results} \label{results}
\subsection{Overall Structure}
The archival image of NGC~839 obtained with the \emph{HST} (proposal 10787, PI: Charlton) shows a galaxy which has a remarkable morphological resemblance to the famous superwind galaxy M82 (Figures \ref{fig1} and \ref{fig13}). It displays a highly inclined irregular disk of stars crossed by aligned filamentary dust lanes and with a nuclear starburst region covering $\sim 1.5$~kpc. Some portions of this nuclear starburst are heavily obscured by the dust lanes, particularly toward the east, giving the impression in low-resolution images that there is more than one nucleus. The images generated from the continuum channels of our data cube show a similar structure, albeit of lower resolution. Our data show a nearly edge on disk galaxy with a similar inclination and P.A. to previous observations (I$=67$\textdegree, PA$=350$\textdegree; cf. ~\citealt{Mendes98}). Images generated in the various emission lines show similar structure to previous Fabry--Perot and narrowband filter observations of the \Ha\ gas. The line-emitting gas extends somewhat asymmetrically above and below the disk by several kpc as traced by the red \Ha emission in Figure 1. 

The \emph{HST} image reveals the details of the starburst nucleus. The central star-forming concentration is very bright and covers some 230~pc. It is surrounded by what appears to be a disk of lower-luminosity clusters, with the outer edge delineated by a complex of dust lanes. Most of the luminosity in the main starburst nucleus is concentrated in a region$\sim 850$pc in diameter.

\begin{figure}
\centering
\includegraphics[scale=0.3]{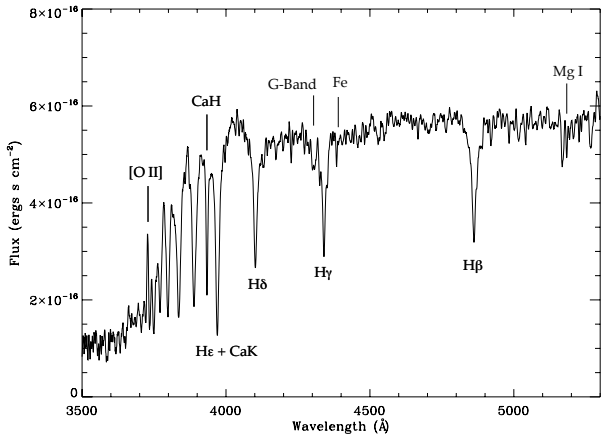}
\caption{Blue WiFeS spectrum extracted from an aperture 3\arcsec. in diameter located in the eastern portion of the disk. The spectrum is presented in rest-frame coordinates. The dominant E+A-type spectrum---which is prominent throughout the stellar disk---clearly shows that the stellar luminosity in the disk of NGC~839 is dominated by an intermediate-age ($\sim 500$~Myr) post-starburst population.}\label{fig3}
\end{figure}

Our WiFeS observations show a strong stellar continuum in the main stellar disk dominated by an A star-like spectrum. An example of this stellar spectrum taken from outside the central line-emitting region is shown in Figure \ref{fig3}. This spectrum, taken by itself, would place NGC~839 in the class of E+A (elliptical plus A-star) galaxies which have undergone a starburst 100 - 1000~Myr ago. However, the spectrum of the core region of the galaxy is clearly that of an active starburst, although a weaker underlying A-type absorption is still seen, suggesting that the galaxy is currently undergoing a transition from a starburst to an E+A galaxy. The stellar continuum displays stronger attenuation in the eastern portion corresponding to an increase in the dust extinction accompanied by a slight increase in electron density $n_{e}$, as traced by the [SII]6717/6730\AA~ratio. Elsewhere in the disk the \SII~ratio provides electron densities on the order of $n_{e}\sim100$ cm$^{-3}$, decreasing to the low density limit ($n_{e}\sim50$ cm$^{-3}$) away from the disk.

\begin{figure}
\centering
\includegraphics[scale=0.3]{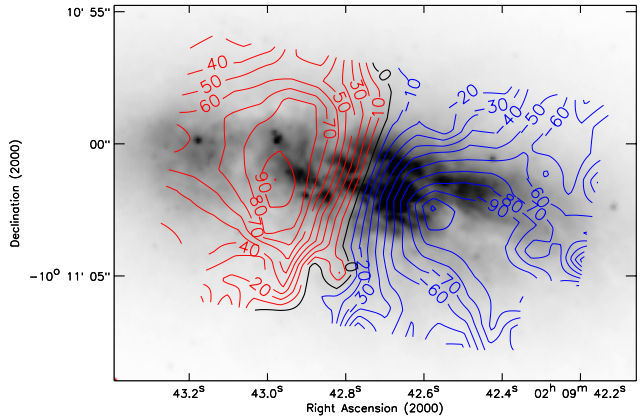}
\caption{Emission line gas rotation map of NGC~839 superimposed on the \emph{HST} ACS F814W archival image. Where the \Ha\ line profile is fit by multiple components, we have used the flux-weighted mean velocity of all components to construct this curve. The rotation curve has not been extended to the polar regions because these are dominated by the outflow motions in the superwind. The maximum velocity of rotation is 100~kms$^{-1}$ and we infer that the enclosed mass in the central $\sim 1.5$kpc is $1.9 \pm 0.2 \times 10^{9}$\Msun. }\label{fig4}
\end{figure}

\begin{figure}
\centering
\includegraphics[scale=0.5]{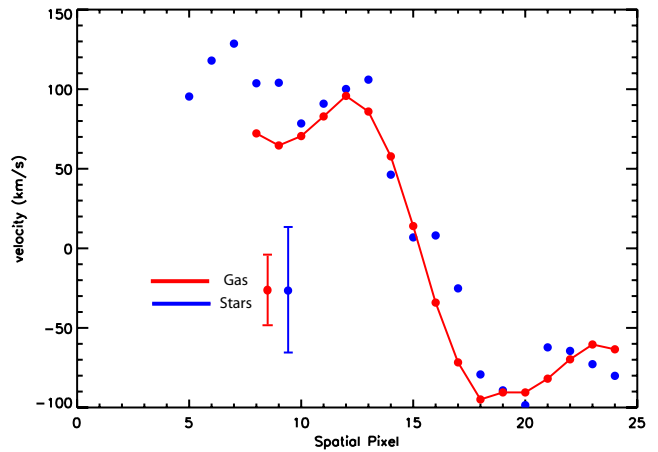}
\caption{Rotation curve generated from the ionized gas velocity (red points) and stellar velocity (blue points).  Values plotted are extracted from a 2\arcsec wide region along the major axis (P.A.$=350$\textdegree). Average error bars for the gas and stellar components are given. Gas velocities are an average of the principal component and stellar velocities are measured from binned spectra. The stellar rotation curve traces the gas rotation within the errors, though there may be an indication of flattening in the eastern part of the disk beyond the ionized gas. Each spatial pixel corresponds to 1\arcsec or 260 parsecs at the distances of NGC 839.}\label{fig5}
\end{figure}

\subsection{Rotation Curve and Mass} \label{rot}
The majority of the emission line gas can be fit by two Gaussians. Along the major axis of the galaxy, we find a gas rotation curve that is more or less well behaved in the inner regions, but which appears to turn down again further out.  Figure \ref{fig4} shows the rotation of the ionized gas, superimposed on the \emph{HST} archive image, the apparent maximum in the gas rotation is coincident with the dust lanes. Figure \ref{fig5} shows rotation curves of the ionized gas traced by emission line velocities and stars traced by Balmer absorption, extracted from a region 2\arcsec wide along the major axis. The stellar data have been averaged from the velocities measured for each spaxel, the stellar data are binned, and the principal component is plotted here. The stellar rotation is more difficult to measure in the starburst region due to weak stellar absorption features, but appears to trace the gas rotation. Beyond the central starburst the stellar rotation may flatten, though it is difficult to discern this given our errors.  The maximum in the gas rotation at a radius of 750~pc corresponds to the boundary of the central starburst region and the enclosed mass $M\times$sin$^{2}$($i$) estimated to lie within this radius is  $1.9\pm 0.2 \times 10^{9}$ \Msun\ . The peak of the rotation curve, at just over 100 km s$^{-1}$, is similar to that found by others \citep{Rubin91,Mendes98}. The maximum rotational velocity is very similar to that found in M~82 by \citet{Shopbell98}. However, the region enclosed is 50\% larger and thus the enclosed mass correspondingly larger. We will make a detailed comparison of NGC~839 with M~82 in Section \ref{M82}.

\begin{figure}
\centering
\includegraphics[scale=0.4]{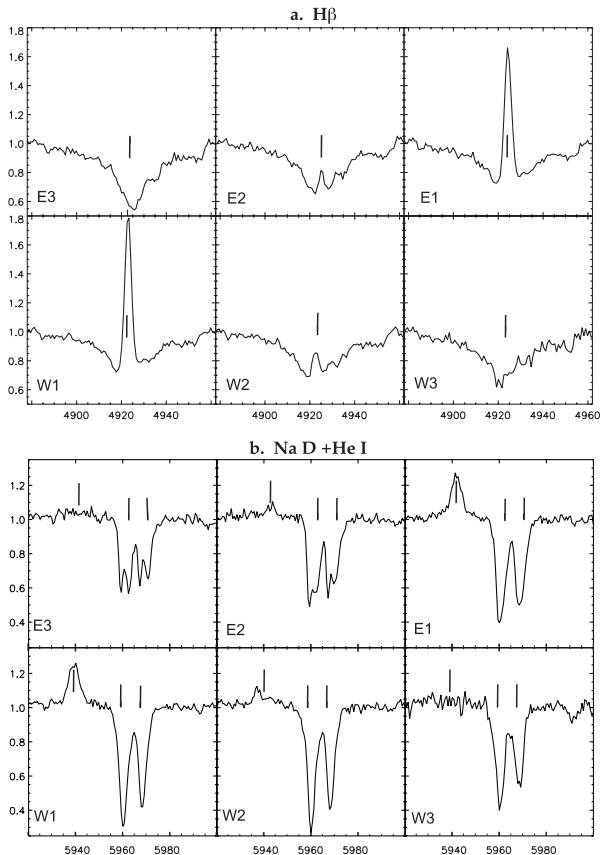}
\caption{Here, we compare (a) the nebular and stellar rotation curve at \Hb\ with (b) the \ion{He}{1}5876\AA~ emission line and much more complex interstellar Na D absorption lines. Each panel integrates a  3\arcsec diameter aperture, and are separated by 3\arcsec each along the major axis of NGC~839. The short bars on each panel give the local systemic velocity of the emission line gas for each spectral feature. Again, the predominance of the A-type stellar absorption is prevalent in the outer regions of the galaxy. Note that the stellar component has a faster rotation velocity than the ionized gas in the outer regions (E3 and W3). The interstellar \ion{Na}{1} lines display a much more complex structure with strong line splitting. One component follows the rotation of the ionized gas, while the other is blueshifted. We infer that this arises in a neutral outflow at the base of the superwind with a projected velocity of $\sim 130$km~s$^{-1}$.}\label{fig6}
\end{figure}

The majority of the emission line gas can be fit by two Gaussians. In the plane of the galaxy, as remarked above, the gas shows an organized rotation curve. This is also true of the stellar component,  shown in Figure \ref{fig6}. Here, we have extracted the spectrum in the region around \Hb\  using several apertures along the major axis of the galaxy each 3\arcsec in diameter. This figure also clearly shows the dominance of the A-type stellar absorption component in the outer parts of the galaxy. The gas appears to have a maximum in its velocity coincident with the dust lanes apparent in the \emph{HST} image (Figures \ref{fig4} and \ref{fig5}).

\begin{figure}
\centering
\includegraphics[scale=0.45]{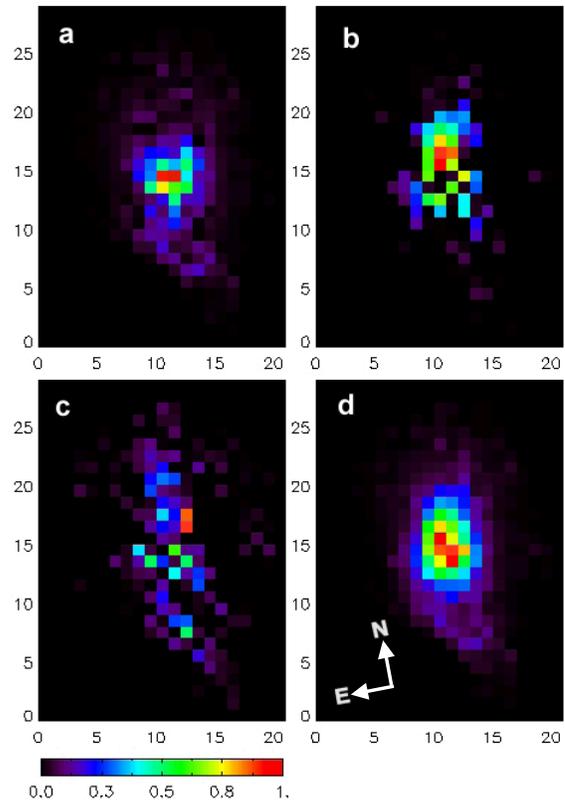}
\caption{Normalized kinetic energy content  $F_{\rm H \alpha} \sigma^{2}$ of the ionized gas along the line of sight, scaled to the peak value. The coordinate scale is in arcsec with the nucleus at $\sim[11,12.5]$, and the analysis has been carried out on individual WiFeS pixels (see Figure \ref{fig1}). The four boxes refer to the energy content of the following components: (a) the energy associated with the velocity dispersion of the principal line component, (b) the energy associated with the velocity dispersion of the secondary component of the line, (c) the energy content associated with the line splitting, and (d) the sum of these three components.  Note how well the outflow axis is defined by component (c), as would be expected for an expanding bicone of gas.}\label{fig7}
\end{figure}

\subsection{Outflow}
The extended \Ha morphology revealed by the SINGG images and by maps generated from our data is typical of that seen in galaxies with biconical galactic scale superwinds (e.g., \citealt{Heckman87,Heckman90,Veilleux94,Lehnert96,Shopbell98,Veilleux02}). The emission extends a few kpc above and below the disk along the minor axis in a conical shape, though due to the distance of NGC~839 and the available ground-based data it is difficult to determine detailed characteristics of the bicone.

Our spectra also reveal kinematic evidence for a large-scale outflow or superwind extending both above and below the disk. We see extensive line splitting with typical $\Delta v \sim 250$ km~s$^{-1}$ along the minor axis up to $\sim2$ kpc above and below the plane. The maximum velocity splitting seen at any point in the flow is $\Delta v \sim 305$ km~s$^{-1}$.

As evidenced by Figure \ref{fig6} there is also evidence for a blueshifted Na D component, which traces the neutral component of the superwind. The largest separation in this NGC~839 is inclined such that we observe the blueshifted component below the disk, along lines of sight toward the disk itself. As the stellar continuum drops rapidly below the disk, we are unable to trace the extent of the Na D much beyond the disk and so focus mainly on the emission line data.

The velocity dispersion of the emission line gas is generally lowest in the disk, on the order of 50 km~s$^{-1}$, and increases above and below the disk to 100--200 km~s$^{-1}$, consistent with the kinematic signatures of a galactic wind (e.g.,~\citealt{Lehnert96}). Where two components are fit the ratio of the velocity dispersions are within a factor of $\sim 1$--$2.5$ of one another. Tracing the velocity dispersion is somewhat complicated due to the presence of multiple Gaussian components of varying relative flux and is limited by our spatial and spectral resolution, so in order to quantify these signatures, we have analyzed the data in terms of the line flux and the square of the velocity dispersion, $\sigma^{2}$, along the line of sight. The \Ha\ line flux is given by $F_{\rm H \alpha} \propto \alpha_{\rm B} n_{\rm H} n_{\rm e} \Delta R$, where $\alpha$ is the effective Case-B recombination coefficient, $n_{\rm H} $ is the hydrogen density in cm$^{-3}$, $n_{\rm e} $ is the electron density in cm$^{-3}$,~and $\Delta R$ is the path length through the ionized medium. This equation can be expressed more simply as  $F_{\rm H \alpha} \propto \Sigma n_{\rm e} $, where $\Sigma$ is the ionized mass surface density. Thus to the extent that the electron density is relatively constant, the \Ha\ flux is a measure of mass surface density. Therefore, the product $F_{\rm H \alpha} \sigma^{2}$ is a measure of the kinetic energy content of the ionized gas along the line of sight. The total energy associated with line splitting is calculated by substituting line width with the deviation of the central peak of each component from the flux-weighted average velocity and summing the product of each components 'deviation with the square of its flux.

In Figure \ref{fig7}, we plot the various components of this energy content, namely, the energy associated with the velocity dispersion of the principal and secondary components (panels (a) and (b)), the energy content associated with the line splitting (panel (c)), and the sum of these three components (panel (d)). In the figure, the energy scale has been normalized to the largest value in the data cube, so as to facilitate comparison between the components. 

The energy associated with the velocity dispersion in each of the components shows a marked tendency to be high within the two funnels of ionized  gas extending almost perpendicular to the stellar disk. This alignment is much more evident in terms of the energy content associated with the line splitting shown in panel (c) and clearly defines an axis for the outflow (P.A.$ = +3$\textdegree $\pm 3$\textdegree). This alignment would be expected if the gas motions lie along the surface of an expanding funnel, since the greatest expansion velocities seen in projection would lie along the axis of the funnel in this geometry. 

Figure  \ref{fig7} clearly shows that we are observing an ionized superwind moving out of the galaxy. The space velocities associated with the outflow are certainly greater than the $\sim 130$km~s$^{-1}$ implied by either the blueshift in the Na D -line absorption or the $\sim 125$km~s$^{-1}$ implied by the line splitting observed in the funnel. The \Ha\ image of Figure \ref{fig1} limits the cone opening angle to not more than 60\textdegree. Therefore, if the predominant gas flow is along the surface of the cone, the actual outflow velocity would be $\sim 250$km~s$^{-1}$.

The outflow at P.A.$ = +3$\textdegree $\pm 3$\textdegree~is neither perfectly aligned along the photometric minor axis of the galaxy or with the rotation axis. The photometric line of nodes measured by \citet{Mendes98} for the stellar disk is at P.A. $= 84$\textdegree $\pm 5$\textdegree. From the HST image, we estimate P.A. $= 80$\textdegree$ \pm 3$\textdegree. Thus, the photometric minor axis is at P.A.$ = 352$\textdegree$ \pm 4$\textdegree. From our rotation curve (see Figure \ref{fig4}), we infer that the polar direction is at P.A.$ = 342$\textdegree$ \pm 3$\textdegree. The mild disagreement between the photometric and kinematic line of nodes had been previously noted by \citet{Mendes03}. This perturbation could be due to the ongoing merger activity in HCG 16.

\subsection{Surface Brightness}
The observations described in the previous section suggest that the geometry of the superwind is close to that of a hollow biconical outflow. Models of superwinds by \citet{Cooper09} show that the optically emitting filaments are entrained and accelerated by a faster and hotter outflow from the nuclear starburst region. In this model, the pressure of the optically emitting material is determined by the ram and thermal pressure of the surrounding hot medium. The filaments are ionized either by the escape of UV photons from the central starburst, or, if the outflowing medium is optically thick to the escape of Lyman continuum photons, then by cloud shocks which propagate into the dense filaments as they are accelerated in the outflow. 

\begin{figure}
\centering
\includegraphics[scale=0.5]{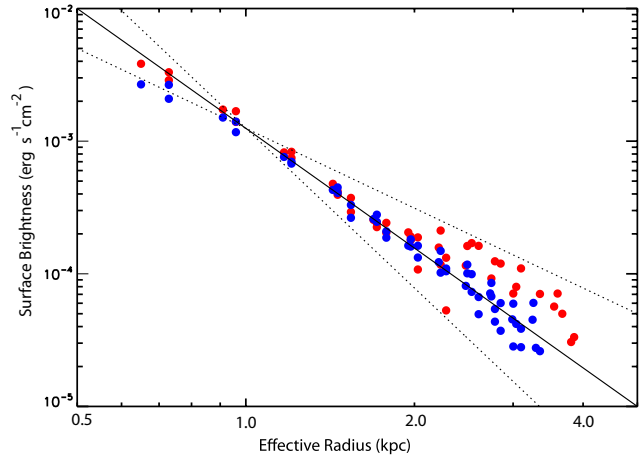}
\caption{H$_{\alpha} $surface brightness as a function of distance ($r$) from the nucleus of NGC~839. The surface brightness of the northern cone is plotted as blue symbols and of the southern cone as red symbols. We have taken a model in which the outflow has the form of two cones with opening angle 30\textdegree meeting at the midplane, and covering the starburst region, so that the starburst is effectively 0.78~kpc from the apex of the cones. The solid line is not a fit, but slope $r^{-3}$. The dotted lines show the shallower $r^{-2}$ and steeper $r^{-4}$ slopes.}
\label{fig8}
\end{figure}

\begin{figure*}[htbp!]
\centering
\includegraphics[scale=0.7]{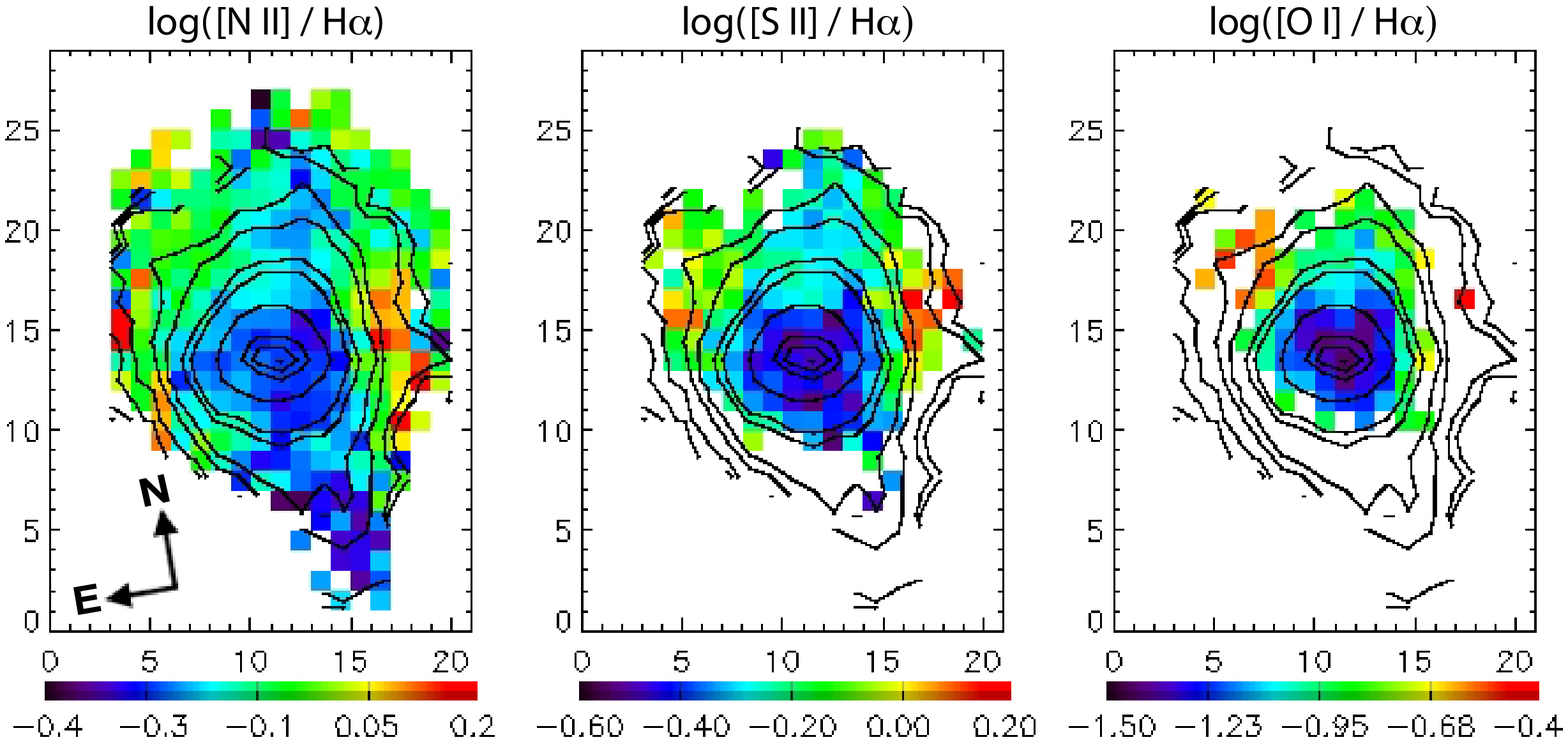}
\caption{Line ratio maps with corresponding diagnostic diagrams for (left to right) log(\NII/\Ha),log(\SII/\Ha), and log(\OI/\Ha). The majority of the emission line gas has line ratios consistent with composite and/or Seyfert/LINER-like emission, and the color bars cover fairly narrow ranges in scale. Overlaid are \Ha flux contours derived from our emission line fluxes. As in Figure \ref{fig7} spatial units are in arcseconds,  with the nucleus at $\sim[11,12.5]$.}
\label{fig9}
\end{figure*}

A study of the radial variation of surface brightness within the cone can help us to distinguish between these possibilities. Let us consider two cases, either (1) that the outflow has constant velocity or (2) that the outflow velocity increases proportional to the radial distance. Recall first the equation derived in the previous section; $F_{\rm H \alpha} \propto \Sigma n_{\rm e} $, where $\Sigma$ is the ionized mass surface density and $n_{\rm e}$ is the electron density. Since the gas pressure is given by $P=nkT$ and the temperature of the ionized plasma is of order $10^4$K, we can write the \Ha\ flux as $F_{\rm H \alpha} \propto \Sigma P $. In a photoionized entrained outflow, the ionized mass is conserved so for constant velocity of expansion, $\Sigma \propto r^{-1}$, where $r$ is the radial distance from the cone apex. The ram pressure in the hot flow decreases as $P \propto r^{-2}$. Therefore, in a constant-velocity photoionized entrained outflow we expect $F_{\rm H \alpha} \propto r^{-3}$. In case (2) (an accelerating outflow), $F_{\rm H \alpha} \propto r^{-4}$.

In a single shocked cloud, the luminosity at \Ha\ is proportional to the energy flux across the cloud shock; $S_{\rm c, H \alpha} \propto A_{\rm c} n_{\rm c} v_{\rm c}^{3}$, where $A_{\rm c}$ is the cross-sectional area of the cloud, $n_{\rm c}$ is the density of the clouds (or filaments), and $v_{\rm c}$ is the cloud shock velocity, given in terms of the external pressure, $P$, by $P = n_{\rm c} v_{\rm c}^{2}$. Therefore, we can rewrite the \Ha\ luminosity produced by a single cloud/ filament as $S_{\rm c, H \alpha} \propto A_{\rm c} P v_{\rm c}$. In a conical outflow, $A \propto r^{2}$, $P \propto r^{-2}$, and $v_{\rm c}$ is constant. Therefore, the flux produced in a single cloud is independent of its distance from the center of the flow. However, the total number of clouds is conserved and their space density decreases as $r^{-2}$ for constant velocity of expansion. In this case, then  $F_{\rm H \alpha} \propto r^{-1}$. In case (2) of an accelerating outflow, $F_{\rm H \alpha} \propto r^{-2}$. In hydrodynamic models by \citet{Cooper09}, the outflowing filaments are shredded and destroyed as they are accelerated, so in this case the radial dependence of the surface brightness could be even steeper than $r^{-2}$.

In Figure \ref{fig8}, we show the variation of surface brightness in \Ha\ along the biconical outflow. We have presented the measurements using a simple bicone with an opening angle of 30\textdegree, which is roughly approximated using our data and the \Ha\ images of \citet{Meurer06}. We assume that the cone originates in the central starburst, which has a radius of $\sim 0.26$ kpc in the disk plane corresponding to the size of the central star-forming concentration seen in the \emph{HST} images. In this case, the starburst lies 0.78~kpc from the origin of the coordinates for each half of the bicone. Figure \ref{fig8} shows that the slope of the surface brightness curve is close to $r^{-3}$, which would tend to support the case for a photoionized outflow within a narrow cone in the polar direction, though this does not rule out the presence of other ionizing sources.

\subsection{Emission Line Ratios}\label{Emrat}
We examine line ratio diagnostic diagrams which use \NII/ \Ha,\SII/\Ha or \OI/\Ha ratios against the  \OIII/\Hb ratio. These diagrams were first proposed by \citet{BPT81} and \citet{VO87} to determine the likely ionizing source of emission line gas in galaxies.  The ionizing source classification on these diagrams was subsequently updated by \citet{Kewley01b,Kewley06} and \citet{Kauffmann03}. These line ratios have the advantage that they can still be used even when reddening corrections are quite uncertain. 

In the case of NGC~839, we find a very systematic behavior in the spatial distribution in the line ratios involving the low ionization species as shown in Figure \ref{fig9}. Figure \ref{fig9} shows that the \NII/ \Ha,\SII / \Ha, and \OI/\Ha ratios are systematically lower near the nucleus, somewhat low along the axis of the bicone, and high along the edges of the bicone. We will compare the observations with detailed models on the optical diagnostic diagrams in the following section.

\section{Analysis} \label{analysis}
Our results of NGC~839 present unequivocal evidence that within the last Gyr it has undergone a long-lived and massive starburst. Not only is the disk dominated by an E+A spectrum originating from this starburst, but the starburst is clearly ongoing and is currently actively powering a superwind evidenced by large bulk motions and line splitting.

The basic physics of superwinds powered by starbursts is rather simple, and has been described by \cite{Chevalier79}. To power a wind, the energy injection per unit area into a fraction, $f$, of the disk gas must exceed its binding energy. As \cite{Chevalier79} have shown, this means that the gas escaping into the halo has an initial temperature greater than some critical value related to the binding energy. Indeed \citet{Cooper09} have re-iterated how similar a superwind is to a simple stellar wind in an inhomogeneous interstellar medium (ISM).

\begin{figure*}[htbp!]
\centering
\includegraphics[scale=0.7]{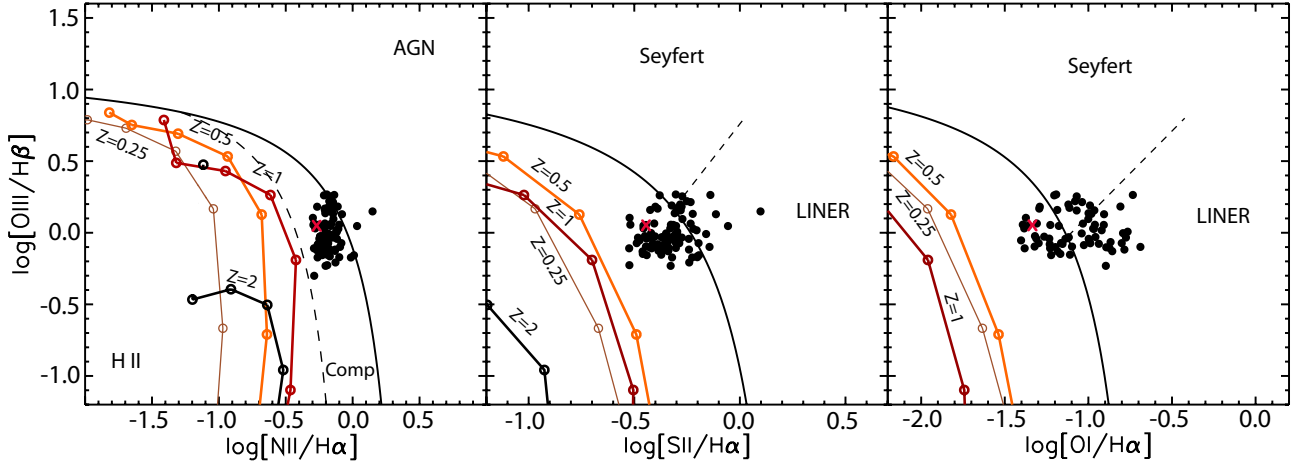}
\caption{BPT diagrams for NGC~839. Filled circles are the line ratios measured in spectra from the individual spatial elements, with our nuclear values marked by a red cross (x). Only data with a minimum of 5$\sigma$ in the relevant diagnostic lines for each diagram are plotted, 91 for \NII/\Ha, 87 for \SII/\Ha\, and 71 for \OI/\Ha. The region of the galaxy from which these data are taken corresponds roughly to the \OI/\Ha map in Figure~\ref{fig9}. We have over-plotted the theoretical values derived from the photoionization models described in the text. The plots also show the earlier maximum starburst envelope as defined by \citet{Kewley01a} and the division between Seyferts and LINERs as defined in~\cite{Kewley06}. }
\label{fig10}
\end{figure*}

The physical parameters which control the production of a galactic wind can easily be derived, as shown by \citet{Dopita08}, and we repeat his argument verbatim here. The input energy $E_{\rm in} $ per unit area is proportional to the surface rate of star formation $ \dot{ \Sigma}_* $ integrated over the lifetime of the starburst, $ \tau_{\rm SB}$;
\begin{equation}
E_{\rm in} = \alpha \dot{ \Sigma}_* \tau_{\rm SB}. \label{Ein}
\end{equation}
The escape energy of the gas is
\begin{equation}
E_{\rm esc} = 0.5 f { \Sigma}_{\rm gas} v_{\rm esc}^2,
\end{equation}
where ${ \Sigma}_{\rm gas} $ is the surface density of gas in the disk. The star formation rate is related to the gas surface density through the \cite{Kennicutt98} star formation law
\begin{equation}
\dot{ \Sigma}_* = \beta{ \Sigma}_{\rm gas}^{1.5}.
\end{equation}
Thus,  a galactic wind becomes possible when
\begin{equation}
{{\tau_{\rm SB} \Sigma_{\rm gas}^{0.5}}\over{v_{\rm esc}^2}} \geqslant  {{f}\over{2\alpha\beta} }\sim \mathrm{const}. 
\end{equation}
Therefore, a wind is favored when the escape velocity is low (such as will be the case in small galaxies), when the starburst continues for a significant period (which is not surprising, since long-lived star formation events contribute more energy to removal of gas), and when the gas surface density is high because the energy input per unit mass of ISM is higher in this case. Our observations clearly demonstrate that NGC~839 fulfills all three of these conditions. Given that shock excitation of the filaments in the superwind is implied by the theoretical models of this type of flow, and also by the observed bulk motions, line ratios, and radial dependence of surface brightness, we clearly need to investigate shock models in more detail. The shock velocities which are suggested by observation (100-250~km~s$^{-1}$) are much lower than the fast shock models described by \citet{Dopita96} and \citet{Allen08}. Indeed, low-velocity shock emission has been a much neglected field since the work of ~\citet{Cox72},~\citet{Dopita76,Dopita77},~\citet{Raymond79}~and \citet{Shull79}.


However, it is also possible to have a superwind which is predominantly photoionized by the escape of UV photons from the central starburst, and this possibility also needs thorough investigation. Indeed the behavior of the line ratios in Figure \ref{fig9} might well be explicable in the context of a photoionization model of the cones, since we would expect to see relatively low-excitation gas toward the nucleus and along the bicone axis, while the higher excitation gas would be found close to the bounding ionization front along the walls of the bicone. Likewise, the slope of the radial dependence of surface brightness given in Figure \ref{fig8} is closer to that expected from photoionization models. In the following subsections, we investigate both shock and photoionization models.

\begin{figure}
\centering
{\includegraphics[scale=0.5]{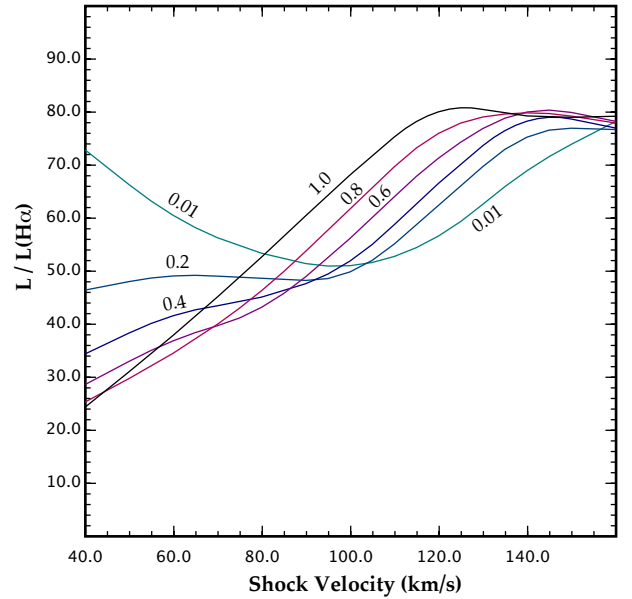}}
\caption{Ratio of total shock luminosity to \Ha\ luminosity as a function of shock velocity and hydrogen pre-ionization fraction as marked on each of the curves.}
\label{fig11}
\end{figure}

\begin{figure*}[htbp!]
\centering
\includegraphics[scale=0.7]{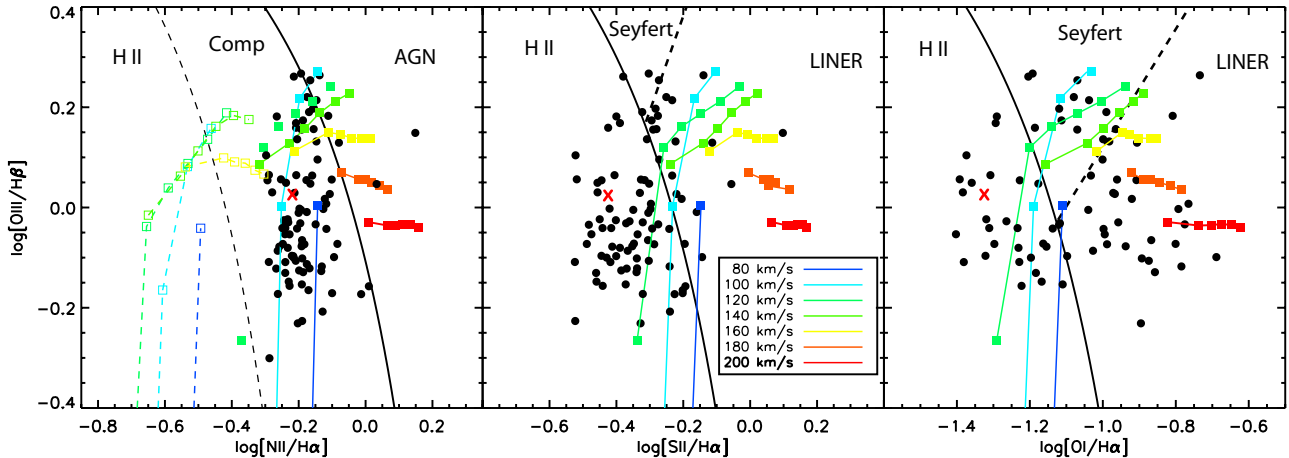}
\caption{This is a blow-up of a region of the BPT diagrams for NGC~839 given in Figure \ref{fig10}. Here, we have overplotted the theoretical values derived from the shock models described in the text. Models plotted are for $Z/Z_{\odot}\sim$2 with the exception of \NII/\Ha, where the dotted lines show the models for solar abundances. Individual points are in order of increasing preionization from bottom to top and left to right, from 0 to 1 in steps of 0.2. The \NII/\Ha plot also shows the maximum starburst (solid black)and composite (dashed black) lines as defined by \citet{Kewley01b,Kauffmann03} and \citet{Kewley06}. The \SII/\Ha and \OI/\Ha plots show the maximum starburst line (solid black) and the division between Seyferts and LINERs (dashed black), as defined in~\cite{Kewley06}. }
\label{fig12}
\end{figure*}

\subsection{Photoionization Models}\label{photo}

In a photoionization model of the superwind, we would expect to see relatively low-excitation gas toward the nucleus and along the bicone axis, while the higher excitation gas would be found close to the bounding ionization front along the walls of the bicone. This can be seen qualitatively in Figure \ref{fig9}. However, this simple model does not stand up to closer scrutiny. As shown in Figure \ref{fig10}, the vast majority of the gas falls within the ``composite" region in the \NII/\Ha diagram, which lies below the theoretical ``maximum starburst line" defined in \citet{Kewley01a} and above the empirical line established in~\citet{Kauffmann03} for pure star formation, conforming to the scheme in \citet{Kewley06}. This implies a radiation field with a harder ionizing flux than can be accounted for by pure star formation. We generated the photoionization models plotted in Figure \ref{fig10} using the STARBURST99 and MAPPINGS III codes \citep{Leitherer99,Sutherland93} with the abundance set of ~\citet{Grevesse10}. Our models are unable to produce hard enough ionizing radiation to reach the theoretical pure star formation lines of \citet{Kewley01a}, which \citet{Levesque10} propose is due to a lack of rotation in the stellar models.

Our observed \SII/\Ha ratios and \OI/\Ha ratios become more \HII region-like toward the nucleus and predominantly LINER-like in other regions. Although \citet{Turner01} suggest the possible presence of a low-luminosity AGN which could possibly account for the \HII/composite line ratios in the nucleus \citep{deCarvalho99}, it cannot account for the extended LINER-like emission. In fact, our line-ratio maps show a general decrease in line ratios closer to the nucleus corresponding to a stronger contribution from star formation, though our optical emission line ratios cannot rule out an obscured AGN and the X-ray observations of \citet{Turner01} and \citet{GonzalezMartin06} provide somewhat ambiguous results. 

\subsection{Shock Modeling}
We have computed a new grid of low-velocity shock models using the MAPPINGS III code, an updated version of the code which was described in \citet{Sutherland93}. In these models, both pre-ionization and shock velocity are treated as independent parameters.  Allowing the pre-ionization to remain a free parameter, rather than being determined by the computed upstream ionizing radiation field is a novel approach which has the advantage of removing geometry dependence and showing the total range of line ratio diagnostic space which can be covered by low-velocity shock models.

The shock velocities range from 60 to 200 km s$^{-1}$ in steps of 20 km s$^{-1}$, with preionization values in the range $0.01-1.0$ in steps of 0.1. The abundance set is taken from~\cite{Grevesse10} and we use standard (solar region) dust depletion factors from \citet{Kimura03}. In using these depletion factors, we are implicitly assuming that the shock is not sufficiently fast to sputter the dust grains advected into the shocked region. The shock models have a transverse magnetic field consistent with equipartition of the thermal and magnetic field ($B=5\mu$G for $n_c=10$cm$^{-3}$; $B\propto n_c^{-1/2}$), and ranged in velocity from 60 to 200~km~s$^{-1}$ in steps of 20~km~s$^{-1}$. Model grids were run using both $Z/Z_{\odot} = 1$ and 2. 

\begin{figure*}[htbp!]
\centering
\includegraphics[scale=1.0]{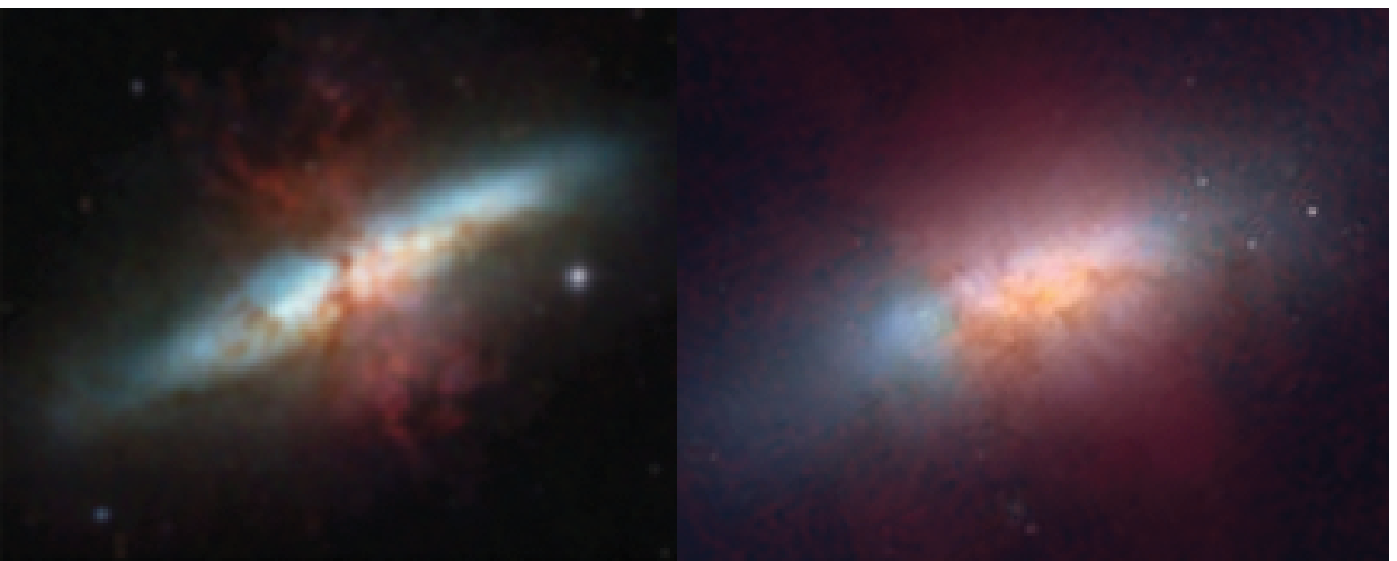}
\caption{Side-by-side comparison of M82 (left) and NGC~839 (right). The image of M82 is an \emph{HST} ACS $B$,$V$,$I$ and \Ha~composite image, the image of NGC~839 is an ACS $B$,$R$ and $I$ image with the \Ha map from SINGG overlaid  ~\citep{Mutchler07,Meurer06}. The resolution has been degraded for the M82 image to provide a better visual comparison with the much more distant NGC~839. NGC~839 has been rotated and scaled  to approximately match the P.A. and spatial scale of M82 to provide a more accurate comparison.}
\label{fig13}
\end{figure*}

For a fully radiative shock propagating at a velocity $v_{s}$ into a cloud  of density $n_{c}$, the surface luminosity $L$ (erg cm$^{-2}$) is equal to the flux of mechanical energy through the shock; $L = \frac{1}{2}n_{c} v_{s}^{3}$.  However, most radiation comes out in UV or IR lines and continuum, and so the optical luminosity does not reflect the shock luminosity. In order to be able to calculate the total shock luminosity, we have computed the ratio of the total to \Ha\ luminosity, $L / L({\rm H \alpha)}$ from our models. This is given for the convenience of others in Figure \ref{fig11}. The ratio $L / L({\rm H \alpha)}$ asymptotes to $\sim 80$ at shock velocities above $\sim 140$~km~s$^{-1}$, and can be approximated as being proportional to the shock velocity for lower velocities;  $L / L({\rm H \alpha)} \sim 60 v_{s} / (100 \rm km~s^{-1})$.

From our grid of shock models we have extracted the \Hb/\OIII, \NII/\Ha,\SII/\Ha and \OI/\Ha line ratios so as to be able to construct the optical diagnostic diagrams consistent with those presented in \ref{Emrat}. These are given in Figure \ref{fig12}. The line ratios derived from the shock models reproduce a significant portion of our observed line ratios from NGC 839. In a broader sense these line ratios provide provide a useful diagnostic tool for understanding the ionizing source in galactic scale winds. Previous studies of galactic  winds suggested the presence of shocks, with shock velocities consistent with our slow shock models (e.g. ~\citealt{Shopbell98,Sharp10}). 

The line ratios for the shock models fall in or very near the composite region in the \NII/\Ha diagram while they migrate into the LINER region in the \SII/\Ha and \OI/\Ha plots. The shock model points plotted in the \OI/\Ha vs \OIII/\Hb plot provide the largest separation in diagnostic diagram space for shock models of different velocities and preionizations. The shock model values for \OI/\Ha also trace the gas with line ratios moving slightly into the Seyfert region of the diagram. Although the separation in \NII/\Ha values derived from the shock models is not as significant as \SII/\Ha or \OI/\Ha, they do provide a larger spread between the models using different metallicities. The shock model values for \NII/\Ha corresponding best to the observed values from NGC~839 are consistent with an abundance of Z/Z$_{\odot}\sim2$, which is similar to the supersolar abundance quoted by~\cite{Turner01}. This can be seen in Figure \ref{fig12}, where the solar abundance models are clearly separate in the \NII/\Ha diagram.

The superwind in NGC~839 exhibits a more composite spectrum than the shock-dominated, starburst-driven winds examined in detail by ~\citet{Sharp10}. In the study by ~citet{Sharp10} the diagnostic diagrams indicate a larger percentage of the extended emission further from the \HII-like region. In this sense the wind in NGC~839 is somewhat more like M82, as the extended emission associated with the wind, while still dominated by shocks, lies nearer the \HII-like region of the diagnostic diagrams. In contrast to M82, however, the extended wind emission in NGC~839 shows a somewhat harder radiation field with higher \OIII/\Hb ratios placing the extended emission in NGC~839 much more firmly in the composite region and beyond the reach of photoionization models as shown in Section \ref{photo}.

\subsection{Comparison of NGC~839 with M82}\label{M82}
NGC~839 shows several similarities to the nearby, well-studied, prototypical wind galaxy M82. A visual comparison reveals similar morphologies, archival \emph{HST} ACS images of both galaxies show an irregular disks crossed by dust lanes, with a strong central starburst (see Figure \ref{fig13}). NGC 839 shows more disk obscuration and less nuclear dust obscuration due to a lower inclination angle. \Ha\ images of both systems show the characteristic biconical outflow, though the filamentary structure of the \Ha\ is easier to discern in M82, as it is 10 times nearer than NGC~839.

Table 1 shows several physical properties of M82 and NGC~839, which provide an even better comparison. The infrared and \Ha\ luminosities (which are associated with the dust and star formation) are quite comparable. The starburst driving the outflows in both galaxies is strongly concentrated around the nucleus. The peak of the gas rotation curve is at a slightly larger radius in NGC 839, though the peak velocities are nearly identical ($\sim100$ km~s$^{-1}$, \citealt{Shopbell98}, and references therein). \citet{Origlia04} measure a gas metallicity in M82 consistent with the metallicity of our shock models, in both cases super solar.

\citet{Shopbell98} provide line ratio maps and diagnostics of M82 that indicate stellar photoionization within the galactic superwind, though they note that the line ratios become more shock-like further from the disk, away from the starbursting region. While the spectra in NGC~839 show a more composite source of ionizing radiation, the line ratios become more \HII region-like close to the nuclear starbursting region. This behavior is consistent with the line ratio structure seen in M82 and the conclusion of \citet{Shopbell98} that close to the nucleus photoionization by the ongoing starburst begins to dominate. Finally, \emph{Chandra} observations of NGC~839 by~\citet{GonzalezMartin06} show funnels of soft X-ray emission along the minor axis similar to those seen  at much higher spatial resolution in \emph{Chandra} images of M82~\citep{Strickland03}. This soft X-ray gas is associated with shocks in the galactic wind and in the case of NGC~839 could be contributing very slightly to the diffuse X-ray emission seen surrounding the entire group by~\citet{Belsole03}.

\begin{deluxetable}{lrr} \label{table1}
\centering
\footnotesize
\tablewidth{0pc}
\tablecolumns{3}
\tablenum{1} 
\tablecaption{Comparison of NGC~839 and M82} 
\tablehead{\colhead{Parameter} & \colhead{M82} & \colhead{NGC~839}}
\startdata
Distance (Mpc)\tablenotemark{a} & 4 & 55 \\
Mass($\odot$)/$R$\tablenotemark{b} (kpc) & $2.3 \times 10^{9}/0.75$ & $2.8 \times 10^{9}/0.5$ \\ 
$L_{H_{\alpha}}\tablenotemark{c}\tablenotemark{d} (ergs~s^{-1})$ & $2 \times 10^{41}$ & $1.25 \times 10^{41}$ \\
$L_{IR} (ergs~s^{-1})\tablenotemark{e}$  & $2.3 \times 10^{44}$ & $3.6 \times 10^{44}$\\
Gas metallicity [O/H] \tablenotemark{b}\tablenotemark{f} & $-3.07$ & $-3.01$ 
\enddata
\tablenotetext{a}{Distances from NED.}
\tablenotetext{b}{Our estimate.}
\tablenotetext{c}{~\citet{Heckman90}.}
\tablenotetext{d}{~\citet{Meurer06}.}
\tablenotetext{e}{~\citet{Sanders03}.}
\tablenotetext{f}{~\citet{Origlia04}.}
\end{deluxetable}

\subsection{E+A Galaxy Connection}
Our observations of NGC~839 suggest a link between lower-mass starburst systems and E+A galaxies.  Although NGC~839 is a fairly low-mass system, it nonetheless has a large-scale organized rotation, and an elliptical-like ($r^{1/4}$) luminosity profile \citep{Mendes98}. The stellar population of the outer regions away from the superwind funnels is dominated by a classical E+A spectrum. These properties are precisely what has been found by \cite{Pracy09} in a Gemini Multi-Object Spectrograph IFU study of a sample of 10 nearby ($z = 0.04-0.20$) E+A galaxies selected from the Two Degree Field Galaxy Redshift Survey. In their sample, they suggest that the large fraction of fast rotators argues against the hypothesis that equal mass mergers are the dominant progenitor to the E+A population. 

NGC~839 and its similarity with M82 suggest a simple scenario for the formation of E+A galaxies. In this, we start with a gas-rich low-mass galaxy in a cluster or loose group. This produces a strong interaction during a close passage with another galaxy. In the case of M82 this would have been M81, while in the case of NGC~839 this was most probably NGC~838, which has a slightly higher IR luminosity ($\log L/L_{\odot} = 11.05$ versus  $\log L/L_{\odot} = 11.01$; ~\citealt{Armus09}), and which also shows a strong nuclear starburst, and evidence of tidal interaction and galactic winds in \Ha\ images. This interaction drives the disk gas toward the nucleus, triggering a strong nuclear starburst. When this occurs, the starburst acts as a source of viscosity in the accretion disk \citep{Collin99}, since shocks generated by increasing mass loss and SN-powered shells reduce rotational angular momentum in the rotationally trailing ISM, and put more angular momentum into the leading ISM, encouraging the development of a galactic wind. The process therefore has the potential to run away, as gas is fed into an ever smaller and denser central region, increasing the specific star formation rate, and driving an ever stronger galactic wind along the lines of Equation (4).

The rotational velocities give further evidence that gas is currently being fed into the central region. As shown above in Section{ \ref{rot}}, the gaseous disk shows a rotational maximum at the boundary of the central starburst region and in the vicinity of the dust lanes seen in the \emph{HST} image. This could be explicable if the outer ionized disk is not in rotational support, and is still infalling into the central star-forming region. A very similar rotation curve was observed in the case of M82 by \citet{Shopbell98}.

\section{Conclusions} \label{conclusions}
In this paper, we have provided very strong evidence for a shock-excited superwind in NGC~839. Our new shock models cannot only reproduce the line ratios in NGC~839, but also the inferred velocities agree with what is derived from emission line and absorption line kinematics. These observations are also consistent with the expected wind-blown shock velocities observed by others \citep{Shopbell98,Veilleux02,Sharp10}, and confirm the model of shock-excited superwinds suggested by recent studies  \citep{Sharp10,Monreal10}. These results are also in line with the picture advocated by \citet{Sharp10} who argue that shocks outshine direct stellar photoionization by the time a galactic scale wind is blown out, but that ongoing star formation contributes to the luminosity and ionization field of the galaxy close to the disk. 

Our shock models clearly demonstrate that galactic-wide emission from intermediate-velocity shocks can shift the global line ratios into the regime of LINERs and ``composite'' objects. Although in the case of NGC~839, the ``composite'' or LINER emission is generated by shocks in a superwind, we can envisage a number of other scenarios which could produce extended LINER-like intermediate velocity shock emission. These include direct galactic collisions,  mergers in which one of the merging galaxies is falling into the stellar wind-blown region of another, and minor mergers involving a gas-rich system falling into the hot gaseous halo of an elliptical galaxy. None of these directly involve an AGN. These results highlight and reinforce the need for integral field data to interpret composite and LINER spectra.

Finally, we have proposed that NGC~839 provides a link between lower-mass starburst systems and the enigmatic E+A galaxies. Outside of the central starburst NGC~839 exhibits a pure E+A spectrum and even within the starbursting nucleus the underlying A-star absorption is still evident. We have proposed a possible formation scenario consistent with the history of NGC~839 given its environment and similarity to M82 which agrees with the recent observational work of ~\citet{Pracy09}. Further integral field observations of systems similar to NGC~839 and M82 should provide further insight into our proposed idea.

\begin{acknowledgements}
We thank the referee for his/her helpful comments, which helped us improve this paper. J. Rich acknowledges support from Spitzer award 5-57359 for the GOALS program. Dopita acknowledges the support from the Australian Department of Science and Education (DEST) Systemic Infrastructure Initiative grant and from an Australian Research Council (ARC) Large Equipment Infrastructure Fund (LIEF) grant LE0775546 which together made possible the construction of the WiFeS instrument. Dopita  thanks the Australian Research Council (ARC) for support under Discovery  project DP0664434. Dopita, Kewley, and Rich acknowledge ARC support under Discovery  project DP0984657. This research has made use of the NASA/IPAC Extragalactic Database (NED) which is operated by  the Jet Propulsion Laboratory, California Institute of Technology, under contract with the National  Aeronautics and Space Administration.  
\end{acknowledgements}

\bibliographystyle{apj}

\end{document}